\documentstyle[12pt]{article}

\begin{document}
\setcounter{page}{1}
\begin{flushright}

{\bf JINR Preprint~~~\\
D1--94--278~~~~~~\\
Dubna, July 1994}\\
\end{flushright}

\begin{center}
{\bf G.I.Smirnov}\\[0.5cm]

A Study of the $A$ dependence of  deep-- inelastic
scattering of leptons and its implications for
understanding of the EMC effect \\[1cm]

Abstract\\[0.5cm]
\end{center}

It is suggested to determine the $A$ dependence of distortions
of the nucleon structure function $F_2(x)$ by summing
 the distortions over an  interval ($x_1, x_2$).

It was found from the analysis of  data on deep-- inelastic
scattering of muons and electrons from nuclei that the
$A$ dependence of  distortion magnitudes obtained
in each of three regions under study, namely
shadowing, antishadowing and the {\em EMC effect} region,
follow the same functional form, being different in the
normalizing factor only. All the available data give
evidence for the saturation of the distortion magnitude
with rising  $A$.\\[2cm]

\begin{center}
Submitted to "Yadernaya Fizika"
\end{center}
\newpage

\begin{center}
{\bf 1. Introduction}\\
\end{center}
\vspace{-0.3cm}

The investigation of distorsions of the nucleon structure
function $F_2(x,Q^2)$ by the nuclear medium, known as the EMC
effect, is a subject of activity even after more
than ten years since the first observation of the effect
from  experiments on the deep-- inelastic scattering
(DIS) of muons on iron and deuterium nuclei.
This is clear from recent publications on the study of
the $A$ dependence of  DIS of electrons in the
E139 -- SLAC experiment \cite{gomez}, from  reports on
the studies of  DIS of muons on nuclei from E665 --
FNAL experiment \cite{carrol} and EMC (CERN) experiment
{}~\cite{ashma}, as well as from  publications on the models
developed for the description of nuclear effects over
the  entire  range of Bjorken $x$ ~\cite{baron,kulag}.
The {\em distorsions} are defined as a deviation from
unity of the ratio $r^A(x) \equiv F_2^A(x)/F_2^D(x)$,
where $F_2^A(x)$ and $F_2^D(x)$ are the structure
functions per nucleon measured on a nucleus of mass $A$
and on a deuteron, respectively.

{}From our point of view the $A$ dependence of nuclear effects
gets much less attention in theoretical papers than the
problem of their $x$ dependence. To a large extent this
is due to the technical difficulties of collecting
the data from DIS experiments on a large number of nuclei.
For instance, the largest number of different nuclear targets
which has been used by SLAC (apart from the deuteron)
is eight but for the range of $x >$ 0.2 only (EMC effect
region). The data from muon DIS experiments in this
kinematic region have poor statistics except for the
BCDMS experiment ~\cite{bari}, which collected data from
two nuclei, N and Fe. In contrast to the E139 -- SLAC
experiment, muon experiments can detect
secondary muons emitted at very small angles (about
2 mrad in the NMC (CERN) experiment), which makes it
possible to study nuclear effects in the shadowing
($x <$ 0.07) and antishadowing (0.07 $ <x<$ 0.2)
regions.

There exists, however, another reason related to the
kind of approach used for the study of the $A$
dependence, namely, all  groups consider  data
at fixed $x$. As a result  $r^A$ either
varies little with $A$ or does not depend on $A$ near
crossover points $x_{I}$, $x_{II}$ and  $x_{III}$, where $r^A$ = 1.
If one  fits  the $A$ dependence with the
relation

$$
r^A = C A^{\alpha (x)}~,
\eqno{(1)}
$$
one obtains~\cite{gomez}  $\alpha (x)$  from
$\sim$ 0 to --0.04 in the interval 0.2 $<x<$ 0.7
and from --0.04 to +0.04 in the interval 0.7$<x<$ 0.9.
The conclusion of this paper is that for each value of $x$
the ratio $\sigma ^A / \sigma^D$ (which is equal to
the ratio $F_2^A / F_2^D$) decreases approximately
logarithmically up to the highest value of $A$ showing no
saturation effects~\cite{gomez}. Similar conclusions follow
from results obtained in a different $x$ range (the
region of nuclear shadowing) in the experiments on
the DIS of muons on nuclei ~\cite{ashma,mallo}.

Despite the evident importance of those results they
are not easy to use in theoretical considerations of the $A$
dependence because of strong correlations of the parameter
$\alpha (x)$ with the $x$ dependence of  $r^A$, which
is not yet
reproduced by any of the suggested theoretical models.
In saying this we of course mean not qualitative agreement
which is provided by the majority of models ~\cite{review}
but a quantitative agreement which can be reached
 in some $x$ intervals only.
One should mention the success of paper ~\cite{baron} in
which  a good quantitative description of $r^A(x)$ has been
reached for the measurements on $~^4$He and deuterium
nuclei. This result has been obtained by considering
nuclear shadowing in the QCD model together with the
hypothesis of nucleon stretching in line with a suggestion
in ref.~\cite{close}. The agreement of the calculations
with the data gets worse when one goes to C and Ca nuclei.

To obtain more informative data on the $A$ dependence
it was suggested in ref.~\cite{sick} to determine
the asymptotic behavior of $\alpha (x)$ which would correspond
to the case of infinite nuclear matter: $A \to \infty$.
The suggestion does not of course eliminate the
correlations. The considered in ref.~\cite{sick}
extrapolation of $r^A$ as a linear function of $A^{-1/3}$
to $A = \infty$ is rather problematic since experimental
errors in each $x$ point do not allow to find
possible deviations from linear dependence.\\

\begin{center}
{\bf 2. Defining the structure function distortion
magnitude as independent of $x$ and $Q^2$}\\
\end{center}
\vspace{-0.3cm}

In contrast to the existing approaches which
consider the $A$ dependence of the DIS of leptons
in terms of the ratio $r^A$ for fixed $x$ value,
in this paper we introduce a conception of the
{\em distortion magnitude} ${\cal M}(A)$ of the
structure function determined from deviations of
$r^A$ from unity in some interval ($x_1, x_2$).
Then the $A$ dependence of $r^A$ is represented as
$$
r^A(x, A) = f(x, {\cal M}) ~,
\eqno{(2)}
$$
where, by the definition, the dimensionless parameter
${\cal M} (A)$ is independent of $x$ and equals
to zero if in the entire interval  $r^A$=1.
This approach exploits the conservation of total nucleon
momentum  carried by partons, from which
it follows that one can not consider the distortion
of the nucleon structure function by the nuclear medium
(as well as the deviation of $r^A$ from unity)
at some point $x$ as
independent of the distortions observed at the adjacent
point $x+ \Delta x$.

As in the conventional approach we consider
structure function distortions as independent of
the 4--momentum transfer $Q^2$ at which $r^A(x)$
is measured. This is justified by  conclusions about the
$Q^2$ independence of $r^A$ in the range  0.2 GeV$^2$
$< Q^2 <$ 200 GeV$^2$ \cite{gomez,bari,review}.\\

Our choice of the function $f(x, {\cal M})$ was
motivated by the possibility to factorize the $x$ dependence
of $r^A$ in the range of 0.003 $< x <$ 0.7 in accordance
with differences in the $r^A$ behavior found in the
three intervals namely the
(1) shadowing, (2) antishadowing and (3) EMC effect
regions:
$$
r^A(x, A) = x^{m_1} (1 + m_2) (1 - m_3 x) ~,
\eqno{(3)}
$$
where the parameters $m_i$ correspond to the introduced
earlier distortion magnitude ${\cal M} (A)$ for each
interval.\\

\begin{center}
{\bf 3. Data analysis in the range
0.003 $\leq x \leq$ 0.7 }\\
\end{center}
\vspace{-0.1cm}

The parameters $m_i$ were determined by fitting $r^A(x)$
measured in the range 0.003 $\leq x \leq$ 0.7 on different
nuclear targets with eqn.(3). Data obtained from the
new generation of  DIS experiments with incident
muons and electrons which have small statistic and
systematic errors were used for the fit along with the
EMC data obtained on C and Ca nuclei in the shadowing
region ~\cite{marne}.
It was required that for each nucleus there should
exist data over the entire $x$ range.\\[0.3cm]
\begin{tabular}{lll}
Nucleus  & Experiment & Number of Points\\
&          & \\
\vspace{-0.3cm}
He  &  NMC  ~\cite{amau} + SLAC  ~\cite{gomez} & 18 + 14  \\
         & &   \\
\vspace{-0.3cm}
C  & EMC  ~\cite{marne} + NMC ~\cite{amau} + SLAC ~\cite{gomez}~   & 9 + 18  +
14 \\
 & &  \\
\vspace{-0.3cm}
Ca  &  EMC  ~\cite{marne} + NMC ~\cite{amau} + SLAC ~\cite{gomez}~   & 9 + 18 +
14 \\
  &  & \\
\vspace{-0.3cm}
Cu  &EMC  ~\cite{ashma}  & 10   \\[0.5cm]
\end{tabular}

We did not include relative normalization of data as
a free parameter although the comparison of the results
on $r^A$ obtained by EMC and NMC in the shadowing region demonstrates
a clear systematic shift.
We used instead the total experimental error determined
by adding statistical and systematic errors at each point
in quadrature. For each of four nuclei  good agreement
($\chi ^2 / d.o.f. \leq$ 1 with eqn.(3) has been found.

The results of the fit are given by the solid lines in fig.1.
The observed agreement is evidence for a universal
form of   eqn.(3) for all nuclei and also for the
increase of the distortion magnitudes $m_i$ with $A$.
The parameters $m_i$ are shown in fig.2 as a function
of $A$. The errors for the $m_3$ values are smaller
than the size of the dots in the plot.
All three groups of points
vary approximately as $A^{1/3}$. By fitting the data with

$$
m_i = k_i A^{1/3}
\eqno{(4)}
$$
\vspace{0.3cm}

\noindent we find that only $m_2$ is in good agreement with
 eqn.(4) : $\chi^2 / d.o.f.$ = 1.16.
The data for all three regions indicate possible
saturation of the $A$ dependence for $A>$ 20.
The observed experimental $A$ dependence is similar
to that  suggested in ref.~\cite{barsh} for the
treatment of structure function distortions within
the framework of the three-- nucleon correlation model.
According to ref.~\cite{barsh} the $A$ dependence of
the EMC effect can be represented by the factor
$\delta (A)$
$$
\delta(A) =
N \Biggl( 1 - \frac{1}{A^{1/3}} -\frac{1.145}{A^{2/3}} + \frac{0.93}{A} +
 \frac{0.88}{A^{4/3}} - \frac{0.59}{A^{5/3}} \Biggr) ~,
\eqno{(5)}
$$
with a normalization constant $N$ = 0.27.
Such an $A$ dependence was obtained by postulating that the nucleons
which reside at the surface of a large nucleus can be
excluded from  consideration because of the sharply
reduced (in accordance with  Woods-Saxon potential shape)
surface density.
The parameter $\delta (A)$ is shown in fig.2 as a dotted line.
Both the data and $\delta (A)$ tend to saturate  at large $A$.
With a mere  change of the normalization constant
in eqn.(5) we obtain three lines (solid lines in fig.2)
which satisfactory follow the fitted parameters $m_i(A)$.

Thus, all the data considered in this section give evidence for
a universal $A$ dependence of the distortion magnitudes
$k_i {\cal M}(A)$ of the nucleon structure function for
all three regions, which is satisfactorily described by
eqn.(5) confirming the expectation of small effects
from the surface nucleons.\\

\begin{center}
{\bf 4. Data analysis in the range 0.2 $\leq x \leq$ 0.7 }
\end{center}
\vspace{-0.3cm}

The part of  eqn.(3) which correspond to the distortions
observed in the EMC effect region coincides (except for the
sign of $m_3$) with the linear dependence used in
 refs.~\cite{emc,bodek,bari}
for quantitative estimation of the EMC
effect in the interval 0.2 $<x<$ 0.6. The straight line
fit was discarded after it had been found that $r^A$
oscillates around $r^A$=1 in the region below $x$=0.3.
Our point is that even this piece of data on $r^A$
considered alone can yield important information on the
$A$ dependence of the structure function distortion
magnitude. With this goal in mind we fitted the data
in the interval 0.2$\leq x \leq$ 0.7
with a straight line:
$$
r^A(x) = a ~- ~ b x ~.
\eqno{(6)}
$$
We used in the fit  the values of $r^A(x)$ obtained in
ref.\cite{gomez} on  He, Be, Al, Ca, Fe, Ag and
Au nuclei along with the data of ref.\cite{bari} (Fe)
and ref.\cite{ashma} (Cu).
The resulting $b(A)$ are plotted in fig.3. Figure 4
 shows the coordinates of the second
crossover point $x_{II}$ obtained as
$$
x_{II} (A) ~ = ~ {{a(A) - 1 } \over {b(A)} }~.
\eqno{(7)}
$$
\vspace{0.3cm}

The solid line in fig.3 corresponds to eqn.(5) with $N=$0.54.
 As seen in  fig.3, the $A$
dependence of the distortion magnitude $b(A)$ is similar to
those observed for $m_i$ plotted in fig.2. In addition one
finds in fig.3 a considerable deviation of the data points
from the $C A^{1/3}$ dependence, which is a natural consequence
of a diffuse surface of  large nuclei. At the same time
eqn.(5) satisfactory describes the data up to  $A\sim$ 64.
With $A$ rising further on  one finds  evidence for
 saturation of the distortion magnitude; the last four
points for $b(A)$ are consistent with $b(A)$ = {\em const} =
0.389 $\pm$ 0.012 ($\chi^2 /d.o.f.$ = 0.6). From our analysis
and also from the data plotted in fig.4 it follows that
 $x_{II}$= {\em const} and
equals 0.273 $\pm$ 0.010 . \footnote{We do not discuss  the $A$
dependence of the two other crossover points $x_I$ and $x_{III}$,
where one needs both higher statistical accuracy and larger number
of nuclei to be able to draw quantitative conclusions}

This result considered with the increase of the
distortion magnitude $b(A)$ (oscillation amplitude) of the nucleon
structure function  observed in fig.3
gives evidence for the simultaneous increase
of distortions, with $A$ from $A$=4 on, in the regions
$x < x_{II}$ and $x > x_{II}$, which is in agreement with the conclusion
on the similarity of the $A$ dependence of $m_i$.

The effect of the $A$ independence of the $x_{II}$ coordinate can
be used for the study of  how parton distributions are distorted by nuclear
medium. This follows from the well known property of  the valence
and sea quarks distributions which reach a maximum on each
 side of the $x_{II}$ point.
 Thus, if the parton distribution of one sort only
were distorted (e.g. that of valence quarks), the coordinate $x_{II}$
would have changed its value with  rising $A$.\\

\begin{center}
{\bf 5. x Dependence of $F_2^{A_1} /F_2^{A_2}$ }\\
\end{center}
\vspace{-0.1cm}

In this section we make use of  eqn.(3) and of the results on the
$A$ dependence obtained in sections {\bf 3} and {\bf 4} to determine
the $x$  dependence of the ratio of structure functions measured  on $A_1$
and $A_2$ nuclei with $A>$ 2.

The form of eqn.(3), the similarity of the $A$ dependence for $m_i$ and
also the saturation observed for $b(A)$ in the $A>$ 20 region allows
one to infer that:\\
\vspace{1cm}

(a) in the region  $x \ll $ 1

$$
F_2^{A_1} / F_2^{A_2} ~=~ C_1 x^{\alpha}~,
\eqno{(8)}
$$
where
$$
\alpha ~=~ m_1(A_1) ~ - ~m_1(A_2)~,
$$
$$
C_1 ~=~ { 1+m_2(A_1)  \over  1+m_2(A_2) }~;
$$
\vspace{0.2cm}

(b) in the range of $x >$ 0.1 and for $A_1  \approx A_2$ or
 $A_1, A_2 >$ 20 the following relation holds:
$$
F_2^{A_1} / F_2^{A_2} \approx ~1 ~.
\eqno{(9)}
$$

We have determined $\alpha^{Ca/C}$ by using  $m_1$ values found
for C and Ca in section {\bf 3} and also $\alpha^{Ca/Li}$ by
interpolating    $m_1(A)$. The ratios
 $F_2^{Ca} / F_2^{C}$ and  $F_2^{Ca} / F_2^{Li}$ determined with
eqn.(8) are shown in fig.5 with solid lines superimposed on
experimental data obtained in ref.\cite{pamau}.
The hatched area is the uncertainty in $\alpha^{Ca/Li}$ which comes
from the error in $m_1(Ca)$ and from the interpolation uncertainty
for the parameter $m_1(Li)$. The data shown in fig.5a favors the
conclusions summarized in eqns.(8)--(9), while the data in fig.5b indicate
a possible discrepancy between eqn.(8) and the experiment.\\

\begin{center}
 {\bf 6. Conclusions}\\
\end{center}

In conclusion, from the analysis of the data on the structure functions
ratio  $F_2^{A}(x) / F_2^{D}(x)$ in the framework  suggested in this
paper,  we have obtained the magnitude of the integral distorsions
of the nucleon structure function in the shadowing, antishadowing and
the EMC effect regions as a function of atomic mass $A$.
The similarity of the $A$ dependence  found in all the three regions makes
it possible to predict the distortion effects in the range $x \ll$1
by using the results obtained in the EMC effect region.

Good agreement is obtained between the data and the function
$r^A(x,A)$ defined by eqn.(3) and an explicit form for the $A$
dependence of the distortion magnitude ${\cal M}(A)$.
The position of the second crossover point $x_{II}$ is found
to be independent of $A$. From these facts, we can draw the
following conclusions:\\

1) The distorsions of nucleon structure functions (or of parton
distributions), that show up as  characteristic oscillations of
 $r^A(x)$ around unity are due to the transition
from $A$=2 to $A$=4; \footnote {We can not make any statement on the
case of a transition from $A$=2 to $A$=3 due to  lack of data
on  $F_2^{^3 He}(x) / F_2^D(x)$}

2)  With the increase from $A$=4 to $\sim$ 20, the shape of the
$x$ dependence of $r^A(x)$ does not change. The distortion
magnitudes (oscillation amplitudes) increase monotonically with $A$ in
 good agreement with the parameter $\delta (A)$~\cite{barsh}
(which serves the purpose of excluding the effect of surface
 nucleons). This increase is similar in all three regions, and
reveals a common $A$ dependence;

3) For $A>$20 the distortion magnitudes
are virtually $A$ independent;

4) The ratios of the structure functions, measured on nuclei
$A_1$ and  $A_2$  with atomic mass $A>2$, are described in the region
$x \ll$ 1 by a simple formula $C_1 x^{\alpha}$, while for the region
$x >$ 0.1 for  nuclei which have nearly equal mass numbers and also for
$A>$ 20, the following relation holds: $~F_2^{A_1} / F_2^{A_2} \approx ~1.$\\
\vspace{-0.3cm}

The conclusions 1) --- 3) give evidence for a unique property of the
4--nucleon system as the system  responsible for the distortions
of the nucleon structure function. The problem of understanding
 the structure of a nucleon bound in a nucleus is thus  split
into explanations of  two phenomena: (a) how the oscillations
turn on for the ratio of $~F_2^{He} / F_2^D$   and (b) how the  distortion
 magnitude ${\cal M}(A)$ ( oscillation amplitude  ) is
amplified with the increase of $A$.

\vfill
\begin{center}
Received by Publishing department\\
      on 8 August 1994.
\end{center}

\newpage
\begin{figure}[h]
\vspace*{14.5cm}
\vfill
{\em Fig.1. The results of the fit with eqn. (3) of the
$F_2^A / F_2^D$ measured on  $He$ ~\cite{gomez,amau},
 $C$ ~\cite{gomez,amau,marne}, $Ca$ ~\cite{gomez,amau,marne}
and $Cu$ ~\cite{ashma} }
\end{figure}
\newpage
\vspace*{12.0cm}
\vfill
\begin{figure}[h]
{\em Fig.2. The nucleon structure function distortion
magnitudes $m_i$ versus $A$
determined in the regions of nuclear shadowing ($i$=1),
antishadowing ($i$=2) and  EMC effect ($i$=3).
The dotted line corresponds to the factor $\delta (A)$ ~\cite{barsh},
and the solid lines were obtained by multiplying $\delta (A)$ with
a normalization factor, different for each x -- range. The results
of the fit with  $k_i A^{1/3} $  are shown with dashed lines.}
\end{figure}
\newpage

\vspace*{6.0cm}
\vfill
\begin{figure}[h]
{\em Fig.3. The nucleon structure function distortion
magnitude versus  $A$ determined in the range
 0.2$ \leq x \leq$ 0.7 with a straight line fit of  $r^A(x)$.
The solid line corresponds to the factor $\delta (A)$ ~\cite{barsh}
with normalization $N$=0.54.
The dashed line shows the result of the  fit of the first four
points of $b(A)$ with $k A^{1/3}$}\\
\vspace{7.5cm}

{\em Fig.4. The  coordinates $x_{II}$ which correspond to
 crossing of the straight line $r^A = a - bx$ with   $r^A =$1.}
\end{figure}
\newpage

\vspace*{6cm}
\hfill (a)

\vspace{8cm}
\hfill (b)
\vfill
\begin{figure}[h]
{\em Fig.5. The results of the measurement~\cite{pamau} of
 $F_2^{Ca} / F_2^{C}$ --- (a) and  $F_2^{Ca} / F_2^{Li}$ --- (b)
compared with $C_1 x^{\alpha}$, where ${\alpha}$
was found from  analysis of the  $F_2^A / F_2^D$  ratios}
\end{figure}

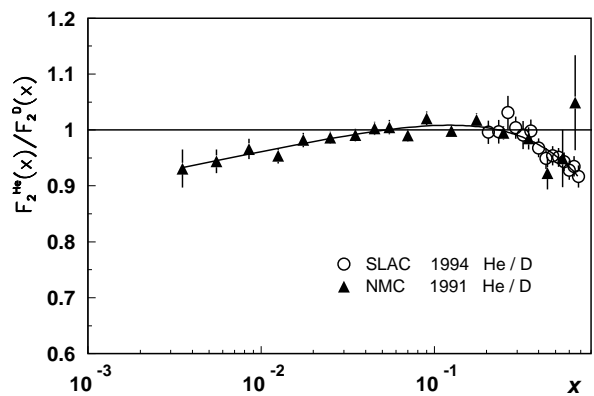
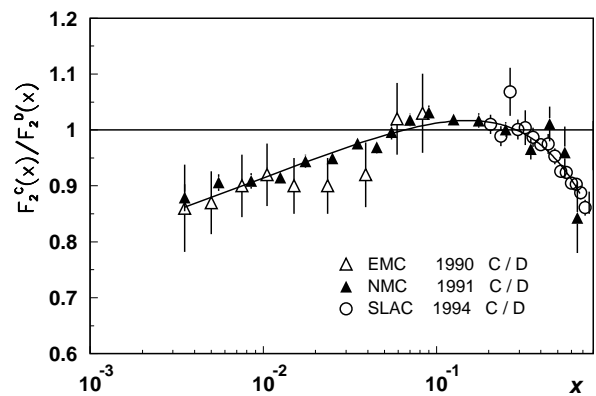
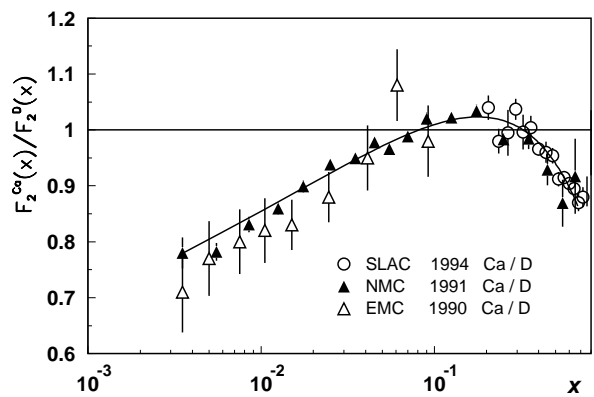
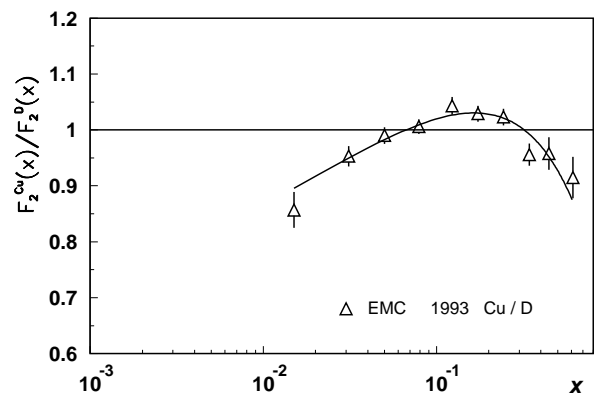
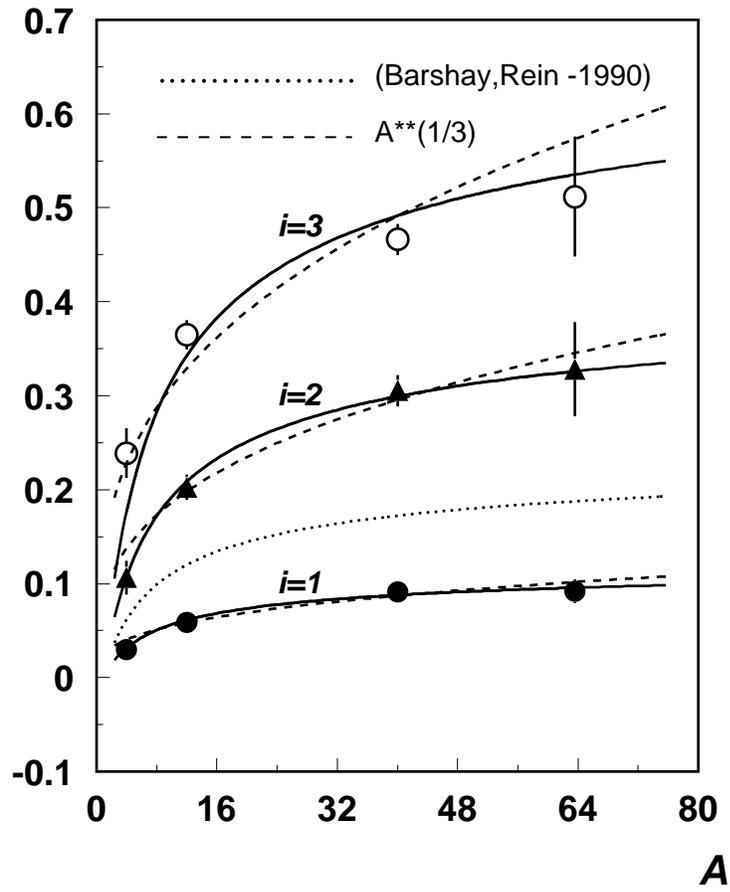
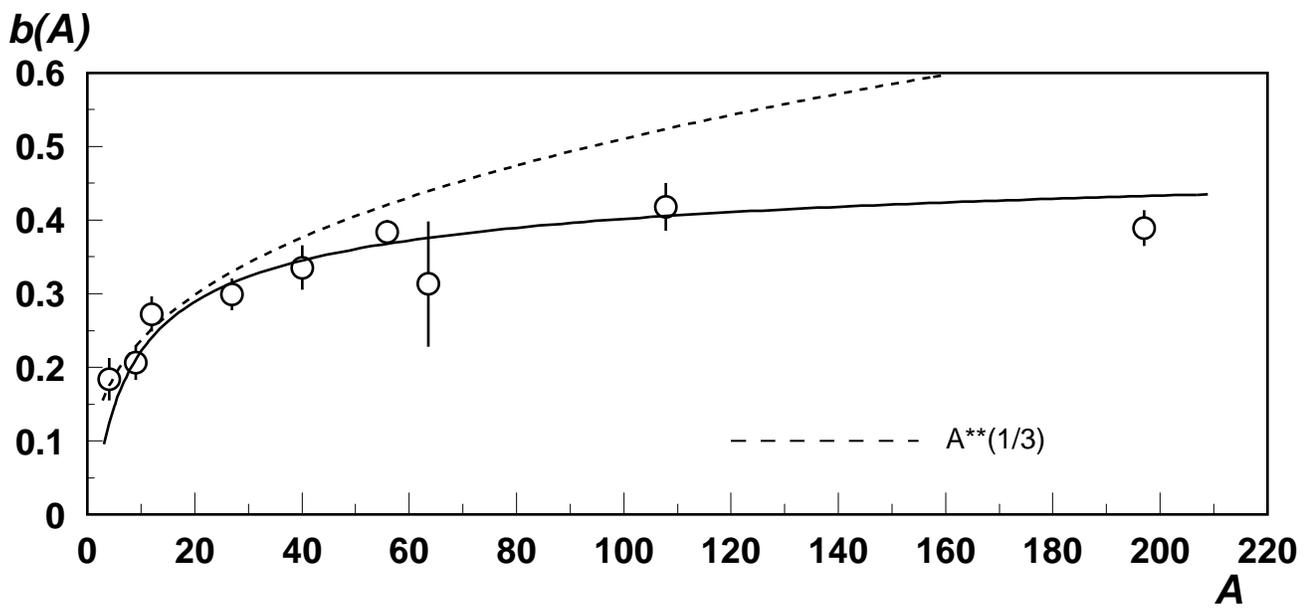
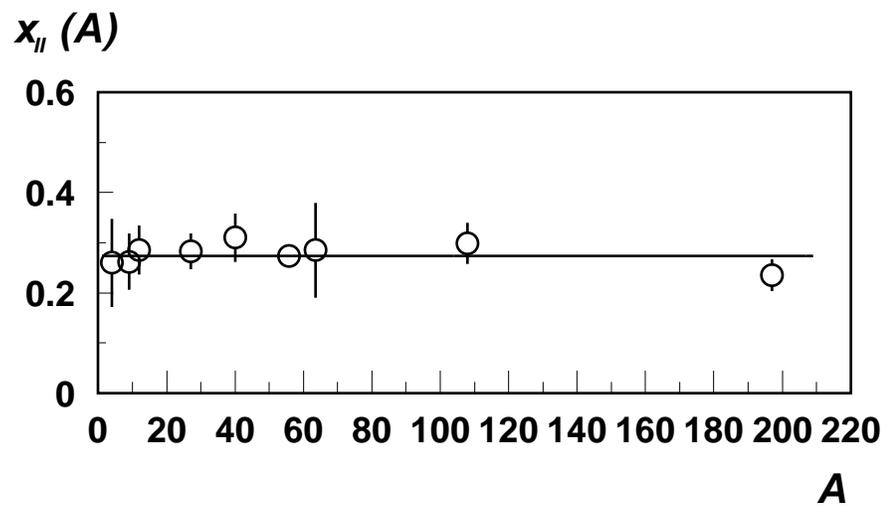
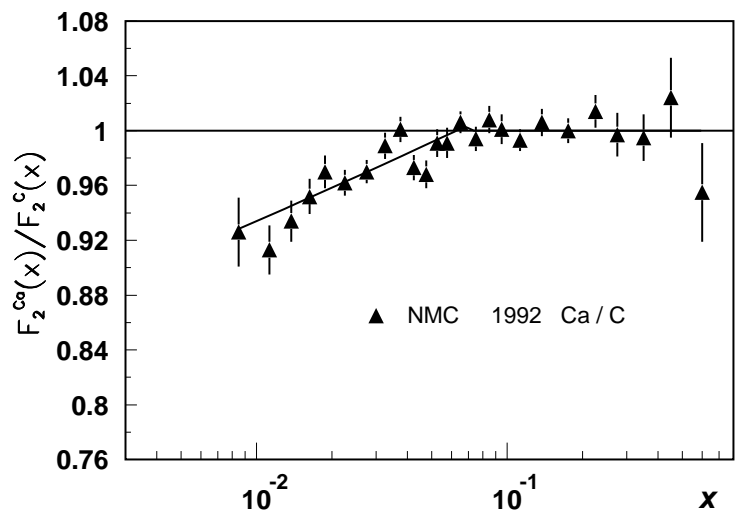
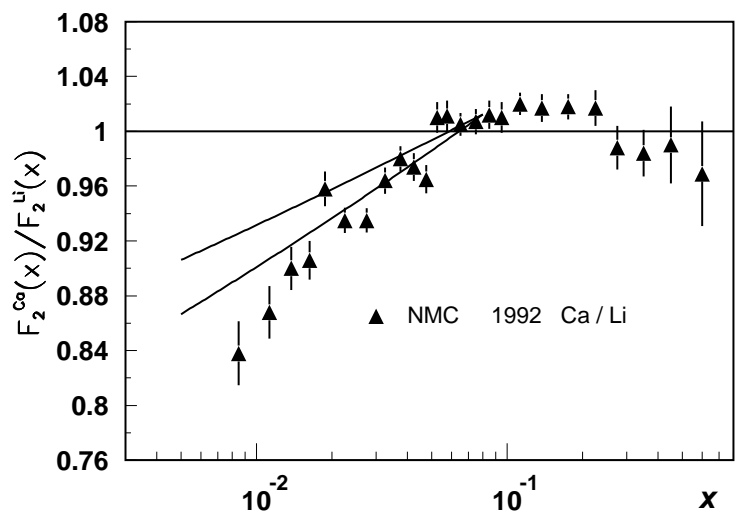
\end{document}